\documentclass{nature}

\usepackage{amssymb}
\usepackage{graphicx}
\usepackage{epsfig}
\usepackage{dcolumn}
\usepackage{bm}
\usepackage{amsmath}

\usepackage{caption}

\usepackage{floatrow}
\usepackage{subfig}

\usepackage{color,soul}

\title{A Memcomputing Pascaline}

\author{Y. V. Pershin$^{1}$, L. K. Castelano$^2$, F. Hartmann$^3$, V. Lopez-Richard $^2$, M. Di Ventra$^4$ }

\begin{document}

\maketitle

\begin{affiliations}
\item Department of Physics and Astronomy, University of South Carolina, Columbia, South Carolina 29208, USA
\item Department of Physics, Federal University at Sao Carlos, Sao Carlos, Sao Paulo 13565-905, Brazil
\item Technische Physik, Physikalisches Institut, Universit\"{a}t W\"{u}rzburg and Wilhelm Conrad R\"{o}ntgen Research Center for Complex Material Systems, Am Hubland, D-97074 W\"{u}rzburg, Germany
\item Department of Physics, University of California, San Diego, La Jolla, California 92093, USA
\end{affiliations}

\begin{abstract}
The original Pascaline was a mechanical calculator able to sum and subtract integers. It encodes
information in the angles of mechanical wheels and through a set of gears, and aided by gravity, could perform the calculations. Here, we show that such a concept can be realized in electronics using memory elements such as memristive systems. By using memristive emulators we have demonstrated experimentally the memcomputing version of the mechanical Pascaline, capable of processing and storing the numerical results in the
multiple levels of each memristive element. Our result is the first experimental demonstration of multi-digit arithmetics with multi-level memory devices that further emphasizes the versatility and potential of memristive systems for future massively-parallel high-density computing architectures.
\end{abstract}

\renewcommand{\thefootnote}{$\star$}

\begin{figure}
  \captionsetup[subfigure]{labelformat=simple,labelfont=bf,font=large}
\begin{center}
  \sidesubfloat[]{\includegraphics[width=10cm]{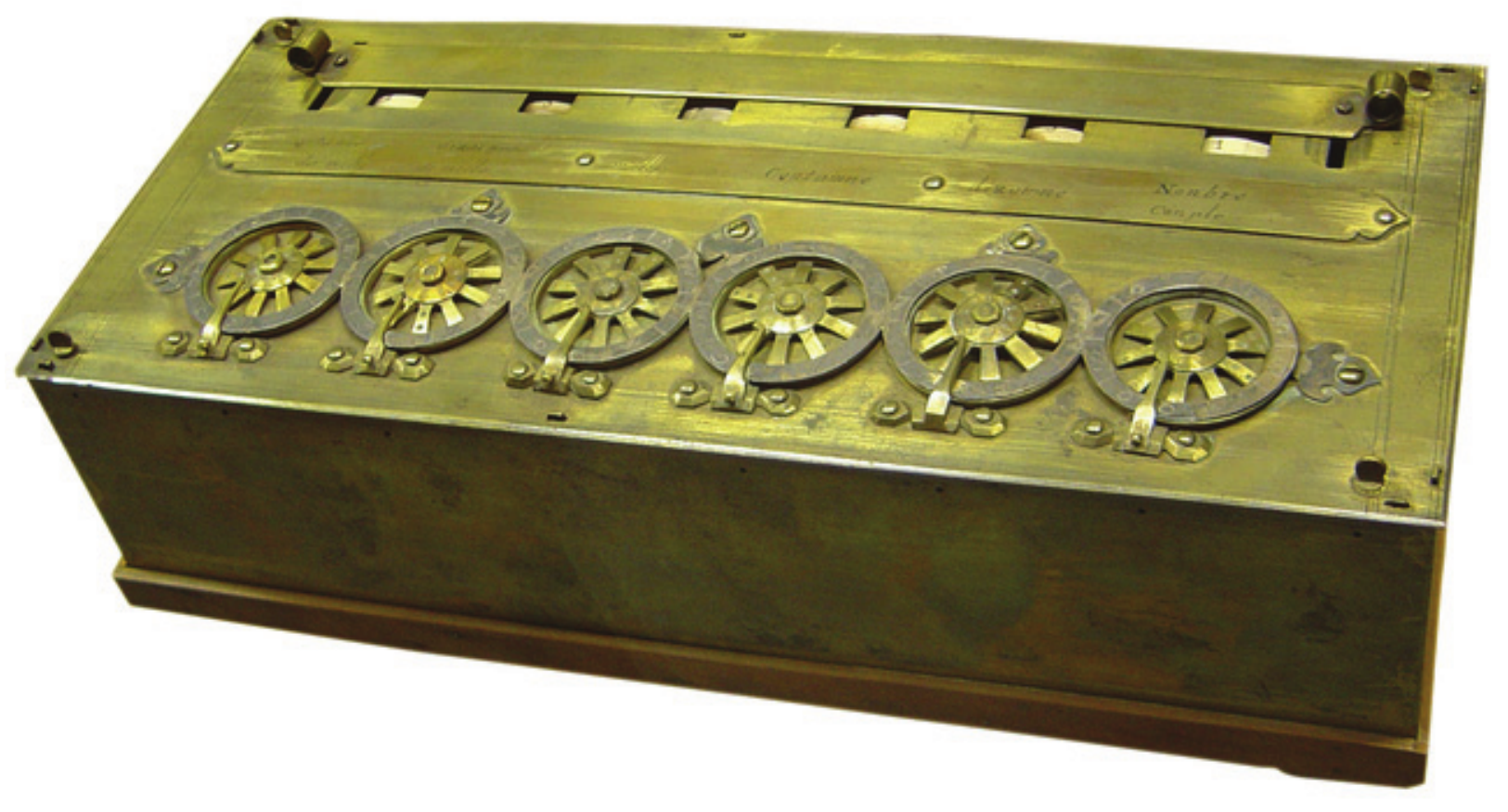}} \\
   \vspace{1cm}
  \sidesubfloat[]{\includegraphics[angle=0,width=15cm]{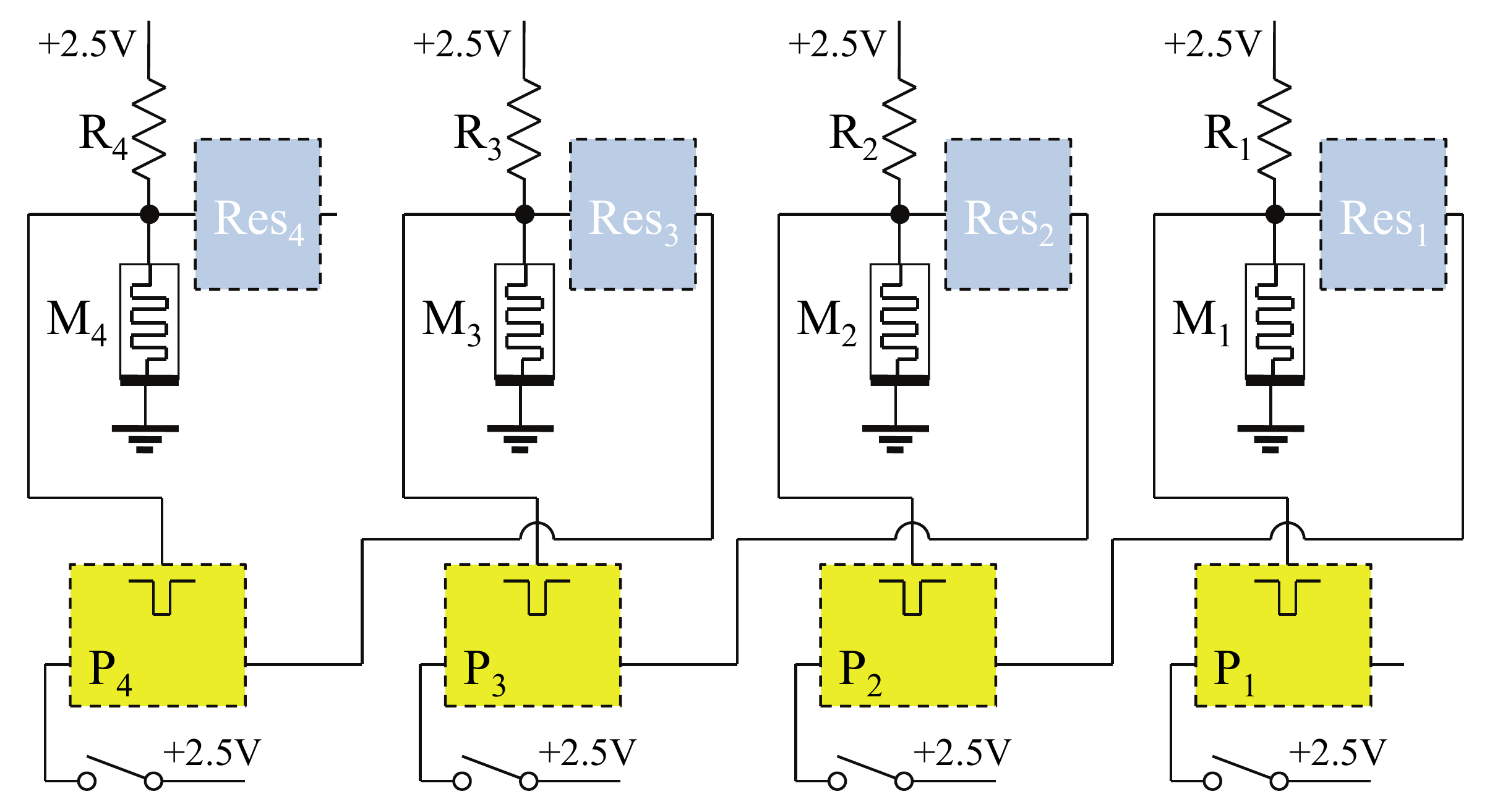}}
\end{center}
\captionsetup{justification=raggedright, singlelinecheck=false}
\caption{{\bf Pascaline machines: from past to present. a.} Mechanical Pascaline (machine de la Reine Christine de Su\`ede, 1652) exhibited in the Mus\'ee des Arts et M\'etiers (Paris).  {\bf b.} Memcomputing Pascaline. Pulse generators P$_1$-P$_4$ (at the bottom) are triggered by push in buttons or carry pulses from Res circuits.  The outputs of P$_1$-P$_4$ (top terminals) provide negative voltage pulses increasing memristances of M$_1$-M$_4$ in steps. Reset circuits Res$_1$-Res$_4$ are used to reset memristors when their memristances exceed a threshold (at the same time generating the carry pulse).}
\label{fig1}
\end{figure}

The Pascaline, built by the mathematician Blaise Pascal, is one of the first known mechanical calculators to perform the basic arithmetic operations \footnote{Practically only addition, since subtraction was performed using the method of nine's complements, multiplication as repeated summation, and division as repeated subtraction.} \cite{pascal1645a} that has been produced on a "large scale" for its times. The machine, whose picture is shown in Fig.~\ref{fig1}{\bf a}, uses gears and wheels and is aided by
gravity to transfer the carry of the addition \cite{pascale1779a}. It encodes ten digits in
certain positions (rotation angles) of the wheels (one wheel for each power of the base-ten system), so we could call it a {\it multi-state machine} in the sense that each of the memory elements (wheels) encodes more than two values of information. In addition, the results of the computation are stored directly in the states (angles) of the wheels themselves, enabling the user to read out the result of the computation without any additional device. Although the commercial success of the Pascaline was limited, it is considered an important step in the development of computing machines.


The Pascaline is of course
far less powerful than our present digital computers. However, these latter ones typically store one bit of information  per memory cell, namely our present digital machines are {\it two-state} (binary) machines. In addition, our present computers employ
a different unit, other than the memory, to actually perform the computation, thus requiring constant communication between the memory and this computing unit. It is thus tempting to employ the multi-state and computing-in-memory features of the  Pascaline in future computing architectures. Indeed, multi-state memory would allow an increased information storage on a single element, and if the results of arithmetic operations on these multiple states could
be stored directly in the memory itself (like for the Pascaline), there would be no need to transfer information from the physical location
where it is computed to where it is stored (at present, this represents an important bottleneck in our digital
computers \cite{Backus78a}).

For the above reasons, the concept of using memory to process information ({\it memcomputing}\cite{diventra13a}) is now receiving increased attention. Its foundations are supported by the mathematical notion of a Universal Memcomputing Machine\cite{traversa14a} -- an alternative computing paradigm to the Turing Machine -- which shows substantial advantages when realized in hardware. Memory elements that store multiple levels of information can be fabricated either using active devices (e.g.,
transistors) or passive ones. The passive ones, in particular, hold the advantage that they require much less power to perform computation than active elements \cite{diventra09a,pershin11a,Chu14a,Kelly14a} and can be realized with a variety of materials and systems down to nanoscale dimensions \cite{pershin11a,diventra11a}.

In this paper, we will demonstrate experimentally a memcomputing machine that works like the Pascaline. We
will use emulators of memristive elements (resistors with memory) we have introduced previously\cite{pershin09d,pershin09c} to build such a device. We will
show that it is capable of addition and subtraction directly in memory. We choose to work in base-10 but our machine can easily work in any base.

We should note that multi-bit information processing and storage with memristive devices has been demonstrated in several experiments~\cite{nian07a,driscoll09b,wright11a,Kim12a,Xu13a,Kelly14a}, and indeed multi-bit arithmetics with memristive devices has already been considered in the literature, but in a considerably simpler architecture than what we show in this work. For instance, in Ref. \citeonline{wright11a} Wright {\it et. al.} demonstrated a single digit base-10 arithmetics with phase-change materials. The authors of Ref. \citeonline{Xu13a} reported a ``memristive abacus'' based on synaptic Ag-Ge-Se devices that can calculate decimal fractions. All this previous work is limited to single-digit operations that are certainly not enough for any practical purpose.

\begin{figure}
  \captionsetup[subfigure]{labelformat=simple,labelfont=bf,font=large}
\begin{center}
  \sidesubfloat[]{\includegraphics[width=8.5cm]{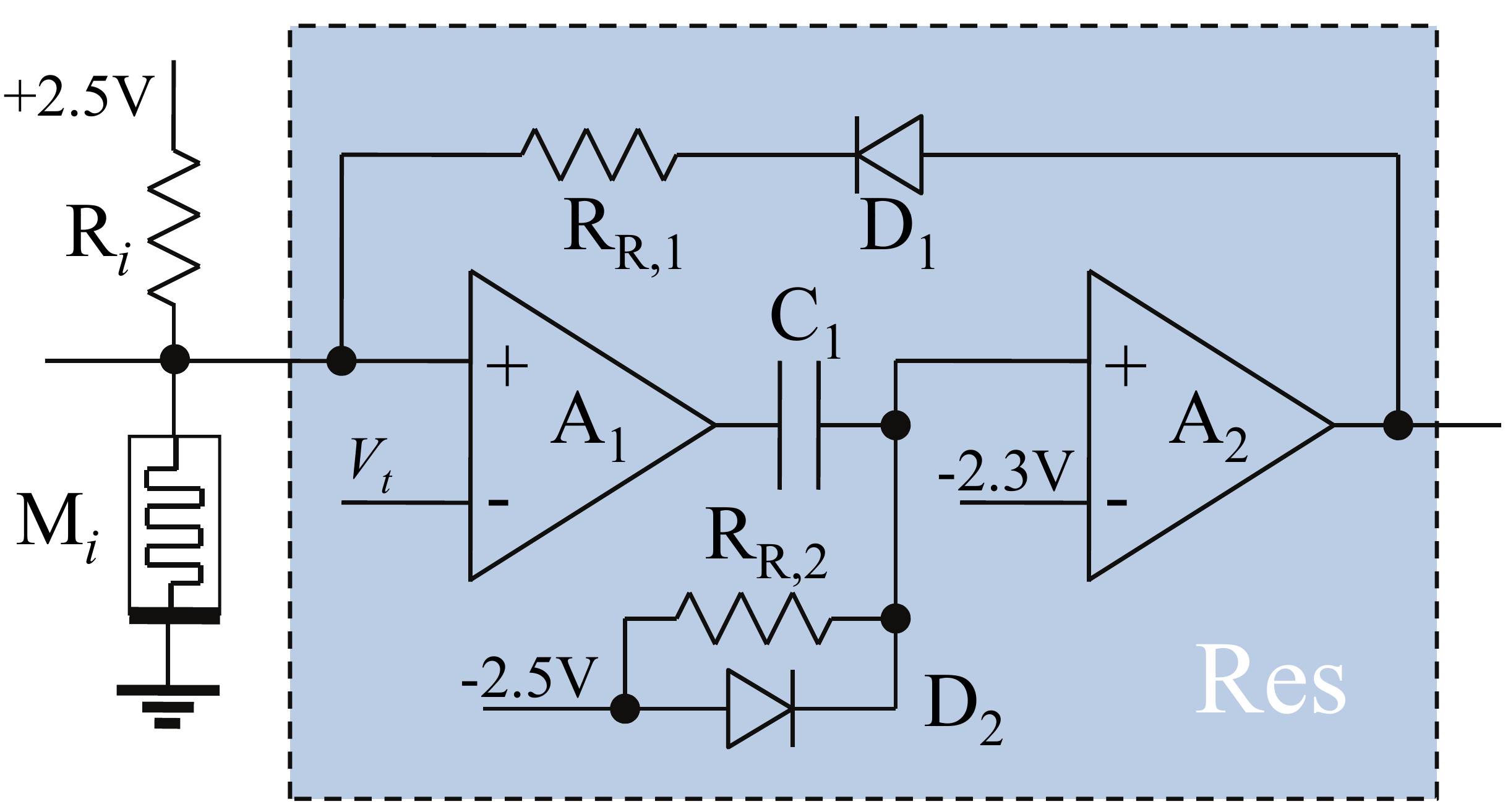}}\quad
  \sidesubfloat[]{\includegraphics[width=6.5cm]{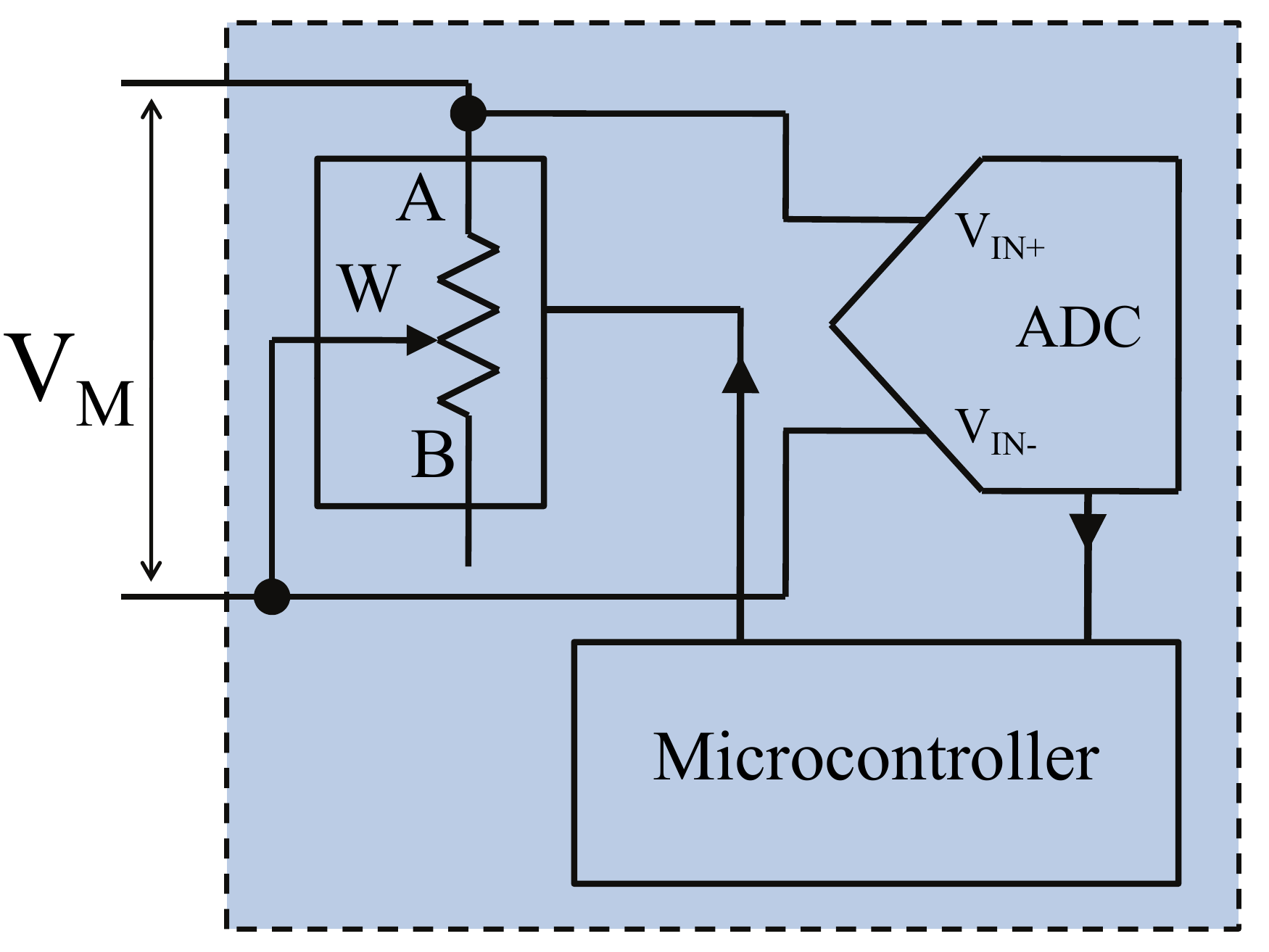}}\\
  \vspace{0.5cm}
  \sidesubfloat[]{\includegraphics[width=7.5cm]{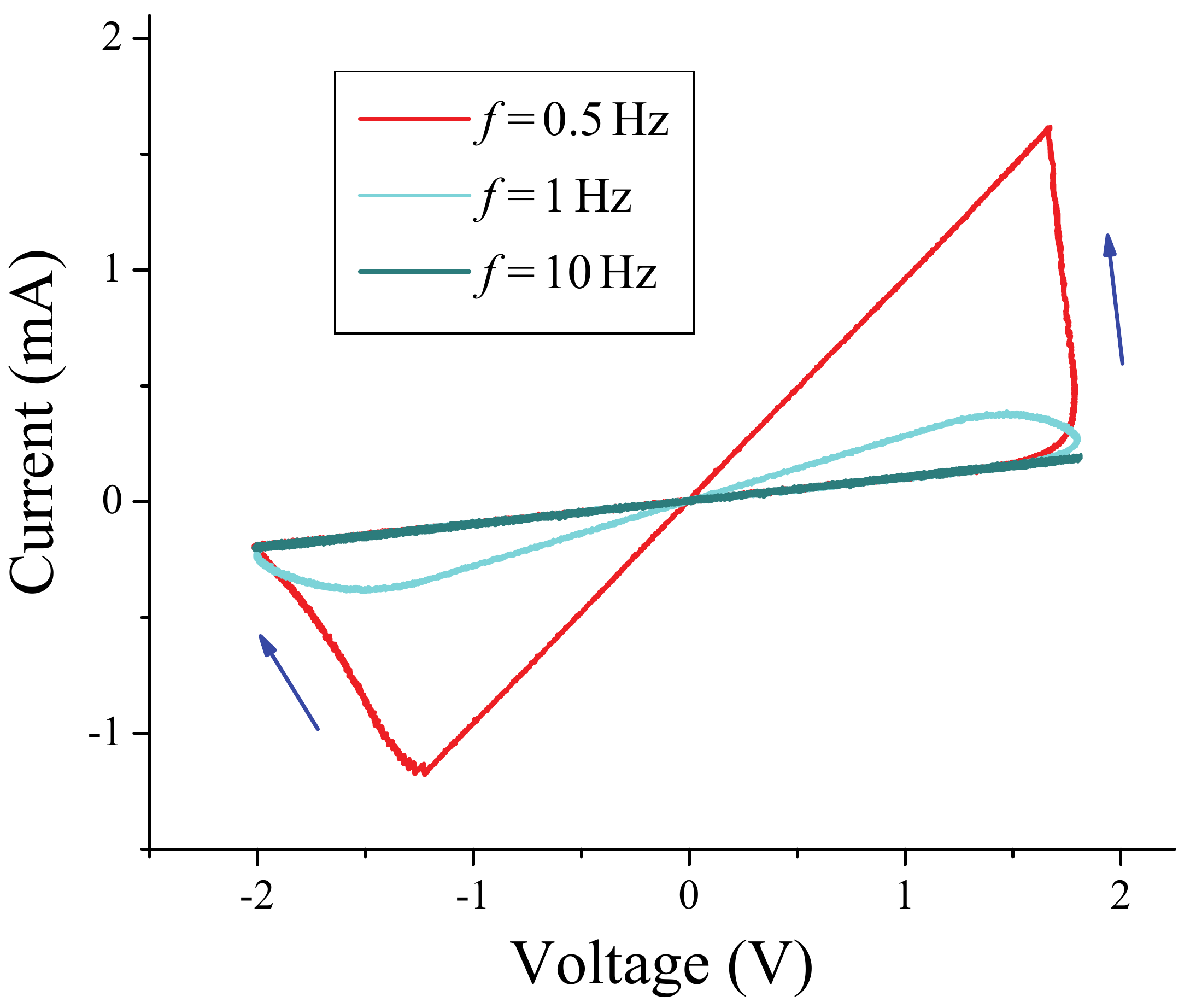}}\quad%
  \sidesubfloat[]{\includegraphics[width=7.5cm]{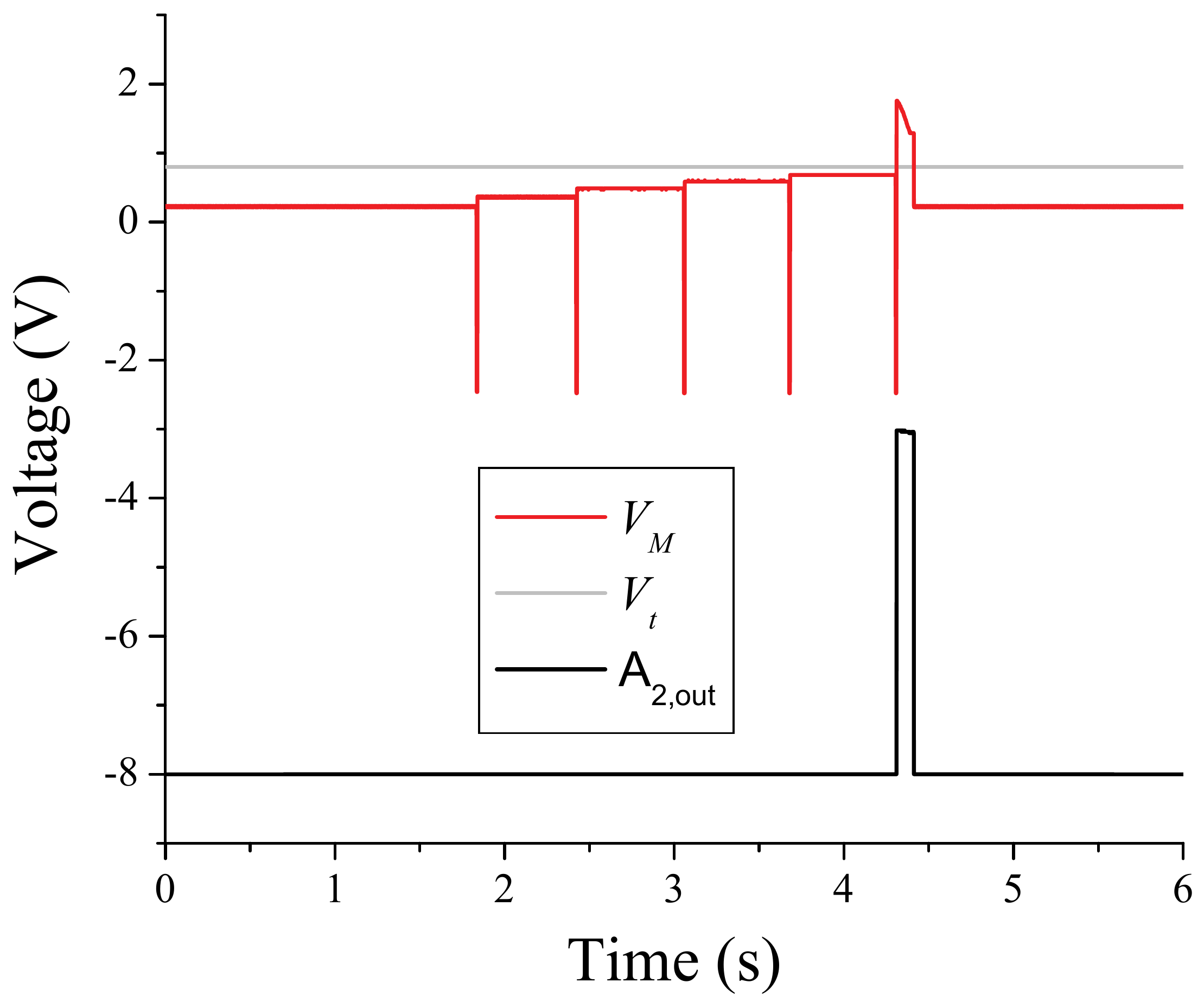}}%
\end{center}
  \caption{{\bf Building blocks of the memristive Pascaline. a.} Schematics of the reset circuit Res$_i$ connected to a memristor. $R_{i}=R_{R,2}=10$ k$\Omega$, $R_{R,1}=470$ $\Omega$, $C_1=10$ $\mu$F, A$_{1(2)}$ are MCP6542 comparators (Microchip), D$_1$ is a common silicon diode, D$_2$ is a low forward voltage germanium diode, and $V_t$ is approximately $0.8$ V. {\bf b.} Schematics of the memristor emulator. The microcontroller continuously updates the resistance of digital potentiometer according to a pre-programmed equation. The applied voltage is measured by the analog-to-digital
converter (ADC).  {\bf c.} Measured $I-V$ curves of memristor driven by an AC voltage source. {\bf d.} Response of the circuit shown in {\bf a} to a train of -2.5 V amplitude, 10 ms width pulses applied to the top terminal of M$_1$. Under these conditions, the circuit implements base-5 arithmetics (note that the memristor resets by the fifth pulse). $V_M$ is the voltage on the memristor, $V_t$ is the threshold voltage, $A_{2,out}$ is the output voltage of the op-amp 2 in {\bf a}.}\label{fig2}
\end{figure}

The experimentally built memristive Pascaline (see Fig.~\ref{fig1}{\bf b}) consists of four identical blocks responsible for four different digits of a number. The blocks are coupled to each other in series so as to carry over. Each block consists of a memristor M$_i$ connected in series with a resistor R$_i$ to $+2.5$ V, a reset circuit Res$_i$, displayed in detail in Fig.~\ref{fig2}{\bf a}, and a pulse generator P$_i$ (here, $i=1,2,3,4$). The digit value is stored in the state of the memristor and represented by the voltage across the memristor in the absence of programming and reset pulses. The memristors are connected in such a way that negative programming pulses from  P$_i$  increase their memristances and, therefore, the voltages across the corresponding M$_i$s. We use threshold-type memristors implemented with memristor emulators \cite{pershin09d,Driscoll10b} (see Fig.~\ref{fig2}{\bf b}). The memristor model \cite{pershin09d} parameters are $\alpha=0$, $\beta=62$ k$\Omega/$V$\cdot$s, $V_T=1.2$ V, $R_{min}=1$ k$\Omega$, $R_{max}=10$ k$\Omega$. Figure \ref{fig2}{\bf c} presents experimentally measured pinched hysteresis loops of a single memristor connected to an AC voltage source. The frequency behavior of these loops is typical for memristive devices~\cite{chua76a}. The memristive Pascaline is powered by a 5V voltage source and employs 2.5V as a virtual ground. The outputs of pulse generators P$_i$ in Fig. \ref{fig1}{\bf c} take $-2.5$ V during the pulse, and are otherwise in a non-connected state.

The reset circuit Res (Fig. \ref{fig2}{\bf a}) compares the voltage across the memristor $V_M$ with a threshold voltage $V_t$ (applied to the 'minus' terminal of A$_1$) and generates a positive (+2.5 V) reset pulse as soon as the voltage on the memristor exceeds $V_t$. The reset pulse width is defined by $R_{R,2}C_1$ time constant (see Fig.~\ref{fig2}{\bf a}) that needs to be sufficiently long to set the memristor into its $R_{min}$ state. This pulse is also used for carry over: as it can be seen from Figure \ref{fig1}{\bf b}, the output of Res$_i$ is connected to the input of P$_{i+1}$.

The present Pascaline architecture supports different number bases, which are defined by the length of the memristor programming pulses (from pulse generators P$_i$) and thresholds of reset circuits $V_t$. For example, using 10 ms programming pulse width, we have obtained a base-5 operation, as demonstrated in Figure \ref{fig2}{\bf d}. The size of the base can be programmed by tuning the pulse width. For instance, the memristor device emulated here, allows a maximum base size determined by $\lceil (R_{max}-R_{min})/(\beta\Delta\tau\Delta V)\rceil$, $\Delta\tau$ being the pulse width, $\Delta V$  the difference between the pulse amplitude and the threshold voltage, $V_T$, and $\lceil$ $\rceil$ denotes the ceiling value. A memristive analog of a French monetary (nondecimal) Pascaline could be realized by programming pulse widths for each digit or selecting different values of $V_t$, that, according to the reset circuit configuration in Figure \ref{fig2}{\bf a}, introduces additional constraints to the size of the base.

\floatsetup[figure]{style=plain,subcapbesideposition=top}


\begin{figure}
  \captionsetup[subfigure]{labelformat=simple,labelfont=bf,font=large}
\begin{center}
  \sidesubfloat[]{\includegraphics[width=10cm]{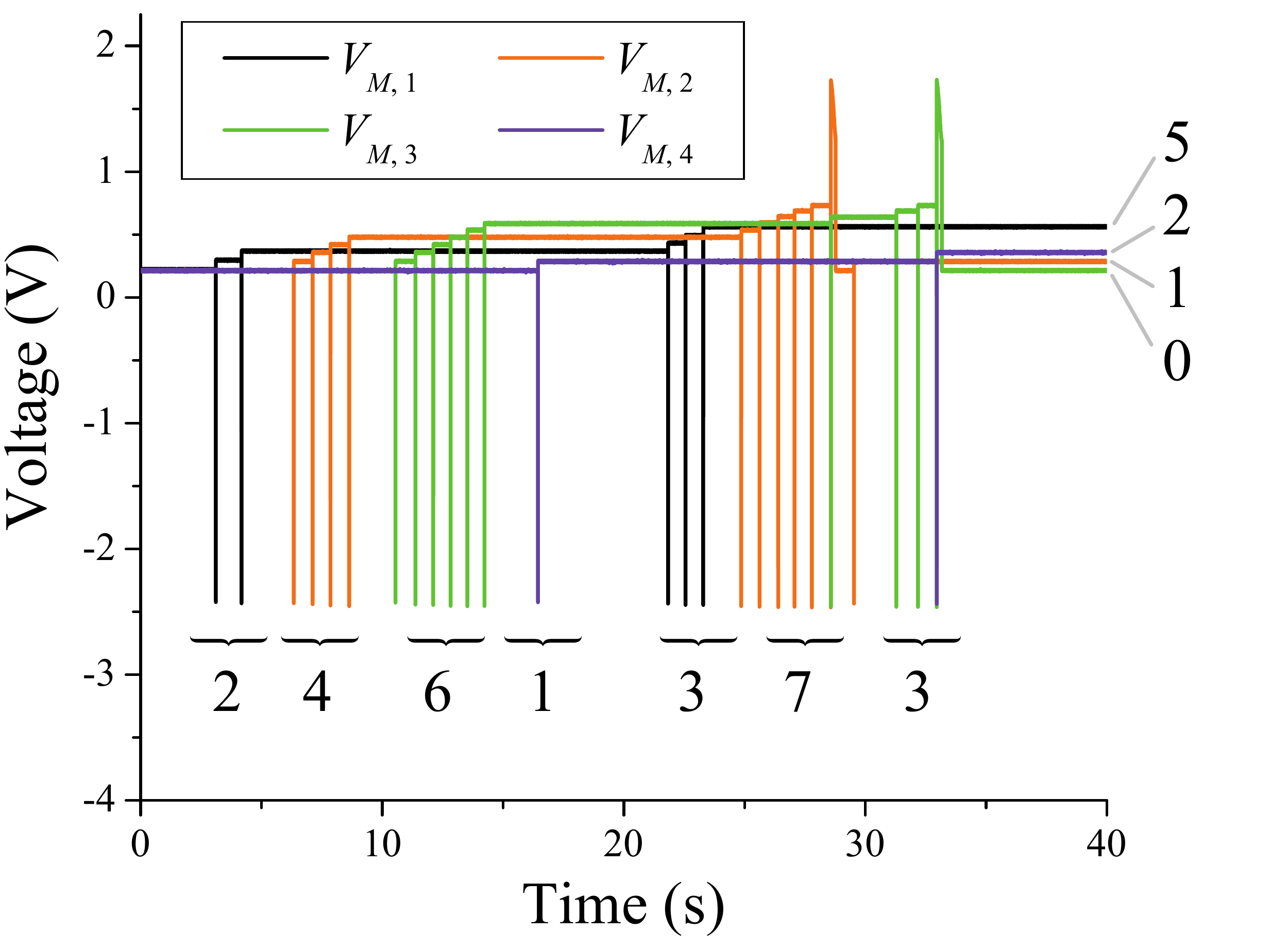}} \\
    \vspace{0.5cm}
  \sidesubfloat[]{\includegraphics[width=10cm]{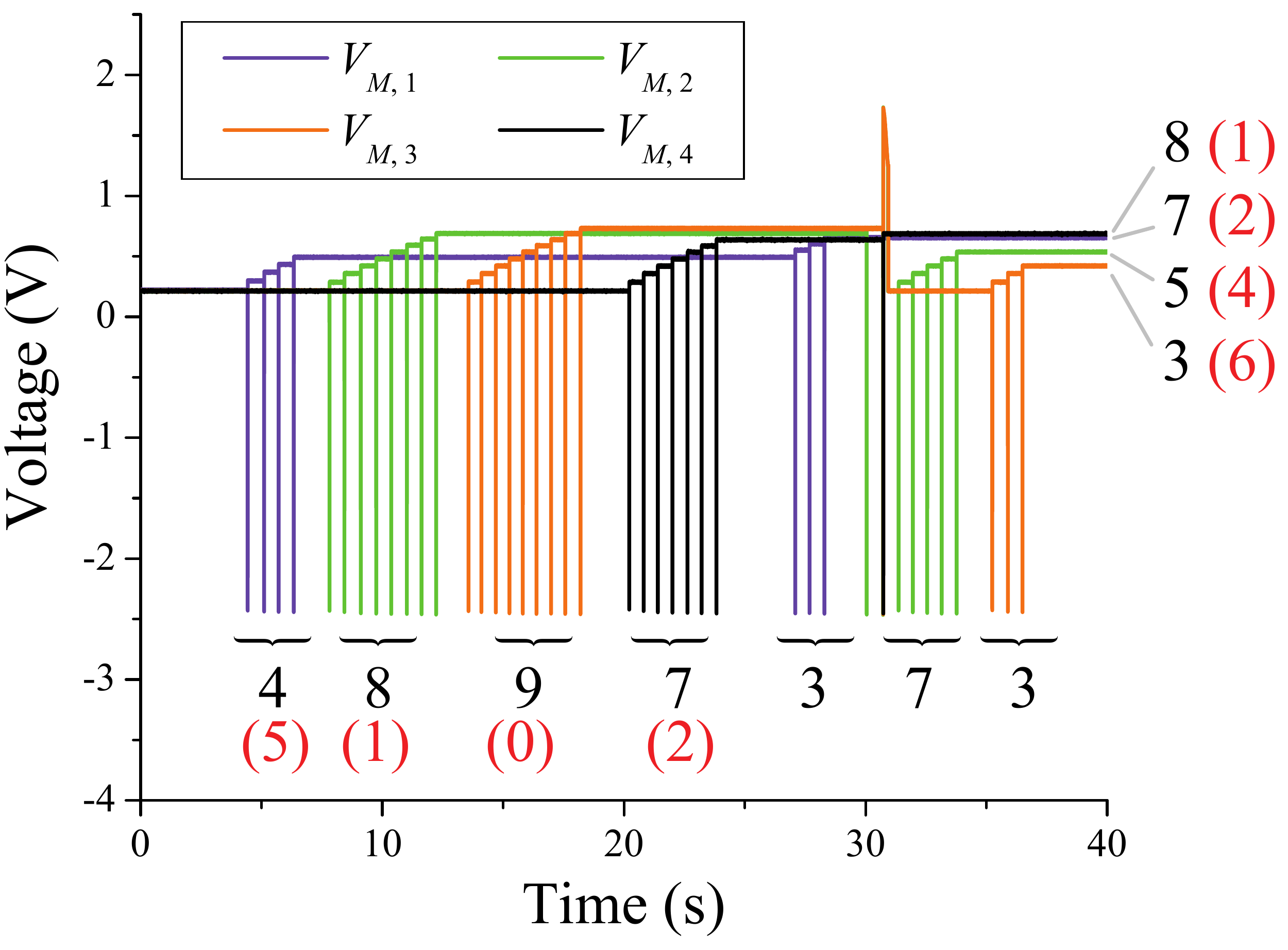}}%
\end{center}
  \caption{{\bf Base-10 addition and subtraction with the memcomputing Pascaline. a.} Addition: $1642+373=2015$. {\bf b.} Subtraction (implemented using nine's complements): $2015-373=1642$. The complements of the first input number and output number are shown in parentheses. $V_{M,i}$ are the voltages on each of the four memristors.}\label{fig3}
\end{figure}

Finally, addition and subtraction with the memcomputing Pascaline is presented in Figure \ref{fig3}. The decimal operation regime has been obtained using 6 ms wide programming pulses. The voltage thresholds, indicated by the digital readouts in Figs. 3 (a) and 3 (b), are determined by the internal parameters of memristors. In the case of the selected device used for emulating the  Pascaline, the resistance levels linearly depend on time and pulse amplitude. As an example of addition, we show that 1642 (the year when the first mechanical calculator was built) plus 373 equals 2015. The subtraction is implemented using nine's complements (the same method is used with the mechanical Pascaline). So, in order to subtract 373 from 2015, we add 373 to the nine's complement of 2015 -- 7984 -- and find the nine's complement of the addition result (8357) -- 1642 --, which is the answer to the subtraction problem.

In summary, we have demonstrated a memcomputing Pascaline that is able to add and subtract using the multiple levels of memory elements (memristors in the present case). Even though we have used emulators of memristors, this machine can be fabricated with actual nano-scale devices and operate also with memcapacitors or meminductors~\cite{diventra09a}, albeit in a slightly different architecture, potentially offering lower energy consumption. The machine can support any type of numerical base, and is the first experimental demonstration of multi-digit arithmetics with multi-level memory devices, an important step forward for future massively-parallel and high-density computing architectures.

\bibliography{memcapacitor}

\begin{thebibliography}{10}
\expandafter\ifx\csname url\endcsname\relax
  \def\url#1{\texttt{#1}}\fi
\expandafter\ifx\csname urlprefix\endcsname\relax\def\urlprefix{URL }\fi
\providecommand{\bibinfo}[2]{#2}
\providecommand{\eprint}[2][]{\url{#2}}

\bibitem{pascal1645a}
\bibinfo{author}{Pascal, B.}
\newblock \bibinfo{title}{Lettre d\'{e}dicatoire \`{a} {M}onseigneur le
  {C}hancelier sur le sujet de la {M}achine nouvellement invent\'{e}e par le
  sieur {B. P.} pour faire toutes sortes d'op\'{e}rations d'arithm\'{e}tique
  par un mouvement r\'{e}gl\'{e} sans plume ni jetons, {A}vec un avis
  n\'{e}cessaire \`{a} ceux qui auront curiosit\'{e} de voir ladite {M}achine
  et s'en servir. {S}uivi du {P}rivil\`{e}ge du {R}oy.} (\bibinfo{year}{1645}).

\bibitem{pascale1779a}
 \emph{\bibinfo{title}{{\OE}uvres de Pascal}} (\bibinfo{address}{La Haye},
  \bibinfo{year}{1779}).

\bibitem{Backus78a}
\bibinfo{author}{Backus, J.}
\newblock \bibinfo{title}{Can programming be liberated from the von neumann
  style? a functional style and its algebra of programs}.
\newblock \emph{\bibinfo{journal}{Comm. Assoc. Comp. Machin.}}
  \textbf{\bibinfo{volume}{21}}, \bibinfo{pages}{613--641}
  (\bibinfo{year}{1978}).

\bibitem{diventra13a}
\bibinfo{author}{Di~Ventra, M.} \& \bibinfo{author}{Pershin, Y.~V.}
\newblock \bibinfo{title}{The parallel approach}.
\newblock \emph{\bibinfo{journal}{Nature Physics}}
  \textbf{\bibinfo{volume}{9}}, \bibinfo{pages}{200} (\bibinfo{year}{2013}).

\bibitem{traversa14a}
\bibinfo{author}{Traversa, F.} \& \bibinfo{author}{Di~Ventra, M.}
\newblock \bibinfo{title}{Universal memcomputing machines}.
\newblock \emph{\bibinfo{journal}{IEEE Trans. Neur. Netw. Learn. Syst. (in
  press)}}  (\bibinfo{year}{2015}).

\bibitem{diventra09a}
\bibinfo{author}{{Di Ventra}, M.}, \bibinfo{author}{Pershin, Y.~V.} \&
  \bibinfo{author}{Chua, L.~O.}
\newblock \bibinfo{title}{Circuit elements with memory: Memristors,
  memcapacitors, and meminductors}.
\newblock \emph{\bibinfo{journal}{Proc. {IEEE}}} \textbf{\bibinfo{volume}{97}},
  \bibinfo{pages}{1717--1724} (\bibinfo{year}{2009}).

\bibitem{pershin11a}
\bibinfo{author}{Pershin, Y.~V.} \& \bibinfo{author}{Di~Ventra, M.}
\newblock \bibinfo{title}{Memory effects in complex materials and nanoscale
  systems}.
\newblock \emph{\bibinfo{journal}{Advances in Physics}}
  \textbf{\bibinfo{volume}{60}}, \bibinfo{pages}{145--227}
  (\bibinfo{year}{2011}).

\bibitem{Chu14a}
\bibinfo{author}{Chu, H.-L.} \emph{et~al.}
\newblock \bibinfo{title}{Programmable redox state of the nickel ion chain in
  dna}.
\newblock \emph{\bibinfo{journal}{Nano Letters}} \textbf{\bibinfo{volume}{14}},
  \bibinfo{pages}{1026--1031} (\bibinfo{year}{2014}).

\bibitem{Kelly14a}
\bibinfo{author}{O'Kelly, C.}, \bibinfo{author}{Fairfield, J.~A.} \&
  \bibinfo{author}{Boland, J.~J.}
\newblock \bibinfo{title}{A single nanoscale junction with programmable
  multilevel memory}.
\newblock \emph{\bibinfo{journal}{ACS Nano}} \textbf{\bibinfo{volume}{8}},
  \bibinfo{pages}{11724--11729} (\bibinfo{year}{2014}).

\bibitem{diventra11a}
\bibinfo{author}{Di~Ventra, M.} \& \bibinfo{author}{Pershin, Y.~V.}
\newblock \bibinfo{title}{Memory materials: a unifying description}.
\newblock \emph{\bibinfo{journal}{Materials Today}}
  \textbf{\bibinfo{volume}{14}}, \bibinfo{pages}{584} (\bibinfo{year}{2011}).

\bibitem{pershin09d}
\bibinfo{author}{Pershin, Y.~V.} \& \bibinfo{author}{{Di Ventra}, M.}
\newblock \bibinfo{title}{Practical approach to programmable analog circuits
  with memristors}.
\newblock \emph{\bibinfo{journal}{{IEEE} Trans. Circ. Syst. I}}
  \textbf{\bibinfo{volume}{57}}, \bibinfo{pages}{1857} (\bibinfo{year}{2010}).

\bibitem{pershin09c}
\bibinfo{author}{Pershin, Y.~V.} \& \bibinfo{author}{{Di Ventra}, M.}
\newblock \bibinfo{title}{Experimental demonstration of associative memory with
  memristive neural networks}.
\newblock \emph{\bibinfo{journal}{Neural {N}etworks}}
  \textbf{\bibinfo{volume}{23}}, \bibinfo{pages}{881} (\bibinfo{year}{2010}).

\bibitem{nian07a}
\bibinfo{author}{Nian, Y.~B.}, \bibinfo{author}{Strozier, J.},
  \bibinfo{author}{Wu, N.~J.}, \bibinfo{author}{Chen, X.} \&
  \bibinfo{author}{Ignatiev, A.}
\newblock \bibinfo{title}{Evidence for an oxygen diffusion model for the
  electric pulse induced resistance change effect in transition-metal oxides}.
\newblock \emph{\bibinfo{journal}{Phys. Rev. Lett.}}
  \textbf{\bibinfo{volume}{98}}, \bibinfo{pages}{146403}
  (\bibinfo{year}{2007}).

\bibitem{driscoll09b}
\bibinfo{author}{Driscoll, T.}, \bibinfo{author}{Kim, H.-T.},
  \bibinfo{author}{Chae, B.~G.}, \bibinfo{author}{{Di Ventra}, M.} \&
  \bibinfo{author}{Basov, D.~N.}
\newblock \bibinfo{title}{Phase-transition driven memristive system}.
\newblock \emph{\bibinfo{journal}{Appl. Phys. Lett.}}
  \textbf{\bibinfo{volume}{95}}, \bibinfo{pages}{043503--1--043503--3}
  (\bibinfo{year}{2009}).

\bibitem{wright11a}
\bibinfo{author}{Wright, C.~D.}, \bibinfo{author}{Liu, Y.},
  \bibinfo{author}{Kohary, K.~I.}, \bibinfo{author}{Aziz, M.~M.} \&
  \bibinfo{author}{Hicken, R.~J.}
\newblock \bibinfo{title}{Arithmetic and biologically-inspired computing using
  phase-change materials}.
\newblock \emph{\bibinfo{journal}{Adv. Mater.}} \textbf{\bibinfo{volume}{23}},
  \bibinfo{pages}{3408} (\bibinfo{year}{2011}).

\bibitem{Kim12a}
\bibinfo{author}{{Kim}, K.-H.} \emph{et~al.}
\newblock \bibinfo{title}{A functional hybrid memristor crossbar-array/cmos
  system for data storage and neuromorphic applications}.
\newblock \emph{\bibinfo{journal}{Nano Letters}} \textbf{\bibinfo{volume}{12}},
  \bibinfo{pages}{389--395} (\bibinfo{year}{2012}).

\bibitem{Xu13a}
\bibinfo{author}{Xu, H.} \emph{et~al.}
\newblock \bibinfo{title}{The chemically driven phase transformation in a
  memristive abacus capable of calculating decimal fractions}.
\newblock \emph{\bibinfo{journal}{Sci. Rep.}} \textbf{\bibinfo{volume}{3}},
  \bibinfo{pages}{1230} (\bibinfo{year}{2013}).

\bibitem{Driscoll10b}
\bibinfo{author}{Driscoll, T.}, \bibinfo{author}{Pershin, Y.~V.},
  \bibinfo{author}{Basov, D.~N.} \& \bibinfo{author}{{Di Ventra}, M.}
\newblock \bibinfo{title}{Chaotic memristor}.
\newblock \emph{\bibinfo{journal}{Applied Physics A (in press)}}
  (\bibinfo{year}{2010}).

\bibitem{chua76a}
\bibinfo{author}{Chua, L.~O.} \& \bibinfo{author}{Kang, S.~M.}
\newblock \bibinfo{title}{Memristive devices and systems}.
\newblock \emph{\bibinfo{journal}{Proceedings of {IEEE}}}
  \textbf{\bibinfo{volume}{64}}, \bibinfo{pages}{209--223}
  (\bibinfo{year}{1976}).

\end{thebibliography}

\begin{addendum}
 \item Y.V.P. was supported by National Science Foundation grant ECCS-1202383. M.D. was supported by CMRR.
 \item[Competing Interests] The authors declare that they have no
competing financial interests.
 \item[Correspondence] Correspondence and requests for materials
should be addressed to Y.V.P.~(email: \\ pershin@physics.sc.edu).
\end{addendum}

\end{document}